\documentclass[aps,prl,twocolumn,preprintnumbers,showpacs]{revtex4-1}
\usepackage{array}
\usepackage{graphicx}
\usepackage{textpos}
\usepackage{hyperref} 

\begin{document}

\title{Algebra of Majorana Doubling}
\preprint{MIT-CTP-4481}

\author{Jaehoon Lee}
\affiliation{Center for Theoretical Physics, Massachusetts Institute of Technology, Cambridge, Massachusetts 02139, USA
}

\author{Frank Wilczek}
\affiliation{Center for Theoretical Physics, Massachusetts Institute of Technology, Cambridge, Massachusetts 02139, USA
}

\date{\today}

\begin{abstract}
Motivated by the problem of identifying Majorana mode operators at junctions, we analyze a basic algebraic structure leading to a doubled spectrum.  For general (nonlinear) interactions the emergent mode creation operator is highly nonlinear in the original effective mode operators, and therefore also in the underlying electron creation and destruction operators.  This phenomenon could open up new possibilities for controlled dynamical manipulation of the modes. We briefly compare and contrast related issues in the Pfaffian quantum Hall state.
\end{abstract}


\maketitle


{\it Introduction.}---The existence of Majorana modes in condensed matter systems~\cite{Mourik2012,JackiwRossi, MooreRead,ReadGreen,AliceaRev,BeenakkerRev} is intrinsically interesting, in that it embodies a qualitatively new and deeply quantum mechanical phenomenon~\cite{Wilczek2009, Wilczek2012}. It is also possible that such modes might have useful applications, particularly in quantum information processing~\cite{Kitaev2006, NayakRMP}.  One feature that makes Majorana modes useful is that they generate a doubled spectrum.   Repeated doubling generates a huge Hilbert space of degenerate states, which is the starting point for possible quantum computational applications. In this Letter, we explore the precise algebraic structure underlying that degeneracy.  

Consider the situation where multiple Majorana modes come together to form a junction, as might occur in a network of superconducting wires that support a nontrivial topological phase. Several experimental groups are developing physical embodiments of Majorana modes, for eventual use in such complex quantum circuits.  (For a useful sampling of very recent activity, see the collection of abstracts from Erice workshop \cite{ericeWorkshop}.)  We consider a fundamental issue that arises in analyzing such circuits. For each odd junction of a circuit,  we identify a remarkably simple, explicit nonlinear operator $\Gamma$ that implements the doubling.  We point out interesting algebraic properties of $\Gamma$ and emphasize its tight connection with fermion parity. We find these results, in their power and simplicity, encouraging for further developments of technology using Majorana wire circuits.   In particular, it should be possible, by systematically incorporating the effects of a very broad class of interactions, to put the analysis of mode transport through trijunctions and Josephson couplings on a more general and rigorous footing. 


{\it Review of Kitaev's wire model.} ---Let us briefly recall the simplest, yet representative, model for such modes, Kitaev's wire segment~\cite{kitaevWire}. We imagine $N$ ordered sites are available to our electrons, so we have creation and destruction operators $a_j^\dagger, a_k$, $1 \leq j, k \leq N$, with $\{ a_j , a_k \} \, = \{ a_j^\dagger , a_k^\dagger \} \, = 0$ and $\{ a_j^\dagger , a_k \} \, = \delta_{jk}$.   The same commutation relations can be expressed using the Hermitian and anti-Hermitian parts of the $a_j$, leading to a Clifford algebra, as follows:
\begin{eqnarray}\label{clifford}
\gamma_{2j-1} ~&=&~ a_j + a_j^\dagger, \nonumber \\
\gamma_{2j} ~&=&~ \frac{a_j - a_j^\dagger}{i}, \nonumber \\
\{ \gamma_k, \gamma_l \} ~&=&~ 2\, \delta_{kl}. 
\end{eqnarray}

Now let us compare the Hamiltonians
\begin{equation}\label{trivialH}
H_0 ~=~ -i \sum\limits_{j=1}^{N} \, \gamma_{2j -1} \gamma_{2j} ,
\end{equation} 
\begin{equation}\label{majoranaModeH}
H_1 ~=~ -i \sum\limits_{j=1}^{N-1} \, \gamma_{2j} \gamma_{2j+1}.
\end{equation}
Since $-i \gamma_{2j-1}\gamma_{2j} = 2 a_j^\dagger a_j - 1$, $H_0$ simply measures the total occupancy.   It is a normal, if unusually trivial, electron Hamiltonian.  

$H_1$ strongly resembles $H_0$ but there are three major differences.   One difference emerges if we reexpress $H_1$ in terms of the $a_j$.   We find that it is local in terms of those variables, in the sense that only neighboring sites are connected, but that in addition to conventional hopping terms of the type $a_j a^\dagger_{j+1}$, we have terms of the type $a_j a_{j+1}$, and their Hermitian conjugates.   The $aa$ type, which we may call superconductive hopping, does not conserve electron number, and is characteristic of a superconducting (pairing) state.    A second difference grows out of a similarity: since the algebra Eq.\,(\ref{clifford}) of the $\gamma_j$ is uniform in $j$, we can interpret the products $\gamma_{2j} \gamma_{2j+1}$ that appear in $H_1$ in the same fashion that we interpret the products $\gamma_{2j-1} \gamma_{2j}$ that appear in $H_0$, that is, as occupancy numbers.   The effective fermions that appear in these numbers, however, are not the original electrons, but mixtures of electrons and holes on neighboring sites.  

The third and most profound difference is that the operators $\gamma_1, \gamma_{2N}$ do not appear at all in $H_1$.   These are the Majorana mode operators.   They commute with the Hamiltonian, square to the identity, and anticommute with each other.   The action of $\gamma_1$ and $\gamma_{2N}$ on the ground state implies a degeneracy of that state, and the corresponding modes have zero energy.   Kitaev \cite{kitaevWire} shows that similar behavior occurs for a family of Hamiltonians allowing continuous variation of microscopic parameters, i.e., for a universality class.  Within that universality class one has Hermitian operators $b_L, b_R$ on the two ends of the wire whose action is exponentially (in $N$) localized and commute with the Hamiltonian up to exponentially small corrections, that satisfy the characteristic relations $b_L^2 = b_R^2 = 1$.    In principle, there is a correction Hamiltonian,
\begin{equation}
H_c ~\propto~ -i b_L b_R,
\end{equation}
that will encourage us to reassemble $b_L, b_R$ into an effective fermion creation-destruction pair and realize $H_c$ as its occupation number.
But for a long wire and weak interactions, we expect the coefficient of $H_c$ to be very small, since the modes excited by $b_L, b_R$ are spatially distant, and for most physical purposes it will be more appropriate to work with the local variables $b_L, b_R$.

%
%

{\it Algebraic Structure.} ---The following considerations will appear more pointed if we
explain their origin in the following little puzzle.   Let us imagine we bring together the ends of three wires supporting Majorana modes $b_1, b_2, b_3$.   Thus we have the algebra 
\begin{equation}\label{3Clifford}
\{ b_j, b_k \} ~=~ 2\delta_{jk}.
\end{equation} 
The $b_j$ do not appear in their separate wire Hamiltonians, but we can expect to have interactions
\begin{equation}\label{3junctionH}
H_{\rm int.} ~=~ -i (\alpha \, b_1 b_2 + \beta \, b_2 b_3 + \gamma \, b_3 b_1 ), 
\end{equation}
which plausibly arise from normal or superconductive interwire electron hopping.   We assume here that the only important couplings among the wires involve the Majorana modes.  This is appropriate if the remaining modes are gapped and the interaction is weak --- for example, if we only include effects of resonant tunneling.   We shall relax this assumption in due course.  

We might expect, heuristically, that the interactions cause two Majorana degrees of freedom to pair up to form a conventional fermion degree of freedom, leaving one Majorana mode behind.

On the other hand, the algebra in Eq.\,(\ref{3Clifford}) can be realized using Pauli $\sigma$ matrices, in the form $b_j = \sigma_j$.  In that realization, we have simply $H \, = \, \alpha \sigma_3 + \beta \, \sigma_1 + \gamma \, \sigma_2$.  But that Hamiltonian has eigenvalues $\pm \sqrt{| \alpha |^2 + |\beta|^2 + |\gamma |^2}$, with neither degeneracy nor zero mode.  In fact a similar problem arises even for ``junctions'' containing a single wire, since we could use $b_R = \sigma_1$ (and $b_L = \sigma_2$).   

The point is that the algebra of Eq.\,(\ref{3Clifford}) is conceptually incomplete.  It does not incorporate relevant implications of electron number parity, or in other words, electron number modulo 2.  For the operator 
\begin{equation}
P ~\equiv~ (-1)^{N_e},
\end{equation}
that implements electron number parity, we should have
\begin{eqnarray}
\label{Psquared} P^2 ~&=&~ 1, \\
\label{PHamiltonian} \left[ P, H_{\rm eff.} \right] ~&=&~ 0, \\
\label{Pb} \{ P, b_j \} ~&=&~ 0. 
\end{eqnarray}
Equation.\,(\ref{Psquared}) follows directly from the motivating definition.  Equation.\,(\ref{PHamiltonian}) reflects the fundamental constraint that electron number modulo 2 is conserved in the theories under consideration, and indeed under very broad --- possibly universal --- conditions.  Equation.\,(\ref{Pb}) reflects, in the context of \cite{kitaevWire}, that the $b_j$ are linear functions of the $a_k, a_l^\dagger$, but is more general.  Indeed, it will persist under any ``dressing'' of the $b_j$ operators induced by interactions that conserve $P$.  Below we will see striking examples of this persistence.

The preceding puzzle can now be addressed. Including the algebra of electron parity operator, we take a concrete realization of operators as $b_1=\sigma_1 \otimes  I$,  $b_2 = \sigma_3 \otimes I$, $b_3 = \sigma_2 \otimes \sigma_1$, and $P=\sigma_2 \otimes \sigma_3$.  This choice represents the algebra  Eqs.\,(\ref{3Clifford}), (\ref{Psquared})-(\ref{Pb}). The Hamiltonian represented in this enlarged space contains doublets at each energy level.  (Related algebraic structures are implicit in \cite{akhmerov}.  See also \cite{modeTransport, modeTransport3D, modeTransportJJA, modeTransportQ1DNet} for more intricate, but model-dependent, constructions.)

%
%
{\it Emergent Majorana modes.} ---Returning to the abstract analysis, consider the special operator
\begin{equation}
\Gamma ~\equiv~ - i b_1 b_2 b_3.
\end{equation}
It satisfies 
\begin{eqnarray}\label{goodGammaProperties}
\label{GammaSquared} \Gamma^2 ~&=&~ 1, \\
\label{Gammab}  \left[ \Gamma, b_j \right] ~&=&~ 0, \\
\label{GammaH} \left[ \Gamma, H_{\rm eff.} \right] ~&=&~ 0, \\
\label{GammaP} \{ \Gamma, P \} ~&=&~ 0. 
\end{eqnarray}
Equationns.\,(\ref{GammaSquared}) and (\ref{Gammab}) follow directly from the definition, while Eq.\,(\ref{GammaH}) follows, given Eq.\,(\ref{Gammab}),  from the requirement that $H_{\rm eff.}$ should contain only terms of even degree in the $b_i$s.  That requirement, in turn, follows from the restriction of the Hamiltonian to terms even under $P$.   Finally Eq.\,(\ref{GammaP}) is a direct consequence of Eq.(\ref{Pb}) and the definition of $\Gamma$.   

This emergent $\Gamma$ has the characteristic properties of a Majorana mode operator: It is Hermitian, squares to one, and has odd electron number parity.   Most crucially, it commutes with the Hamiltonian, but is not a function of the Hamiltonian.   We can highlight the relevant structure by going to a basis where $H$ and $P$ are both diagonal.   Then from Eq.\,(\ref{GammaP}) we see that $\Gamma$ takes states with $P = \pm 1$ into states of the same energy with $P = \mp 1$.    This doubling applies to all energy eigenstates, not only the ground state.   It is reminiscent of, but differs from, Kramers doubling.  (No antiunitary operation appears, nor is $T$ symmetry assumed.)

One also has a linear operator 
\begin{equation}\label{linearOp}
w ~\equiv~ \, \alpha \, b_3 + \beta \, b_1 + \gamma \, b_2 \, 
\end{equation}
that commutes with the Hamiltonian.  
However, it is not independent of $\Gamma$, since we have
\begin{equation}
w ~=~   H \, \Gamma.
\end{equation}

The same considerations apply to a junction supporting any odd number $p$ of Majorana mode operators, with 
\begin{equation}
\Gamma ~\equiv~ i^{ p(p-1)/2} \, \prod\limits_{j=1}^p \, \gamma_j.
\end{equation}
For even $p$, however, we get a commutator instead of an anticommutator in Eq.(\ref{GammaP}), and the doubling construction fails.   For odd $p \geq 5$ generally there is no linear operator, analogous to the $w$ of Eq.\,(\ref{linearOp}) that commutes with $H$. [If the Hamiltonian is quadratic, the existence of a linear zero mode follows from simple linear algebra --- namely, the existence of a zero eigenvalue of an odd-dimensional antisymmetric matrix, as discussed in earlier analyses.  But for more complex, realistic Hamiltonians, including nearby electron modes as envisaged below, that argument is insufficient, even for $p=3$. The emergent operator $\Gamma$, on the other hand, always commutes with the Hamiltonian [Eq. (\ref{GammaH})], even allowing for higher order contributions such as quartic or higher polynomials in the $b_i$s.]

Now let us revisit the approximation of keeping only the interactions of the Majorana modes from the separate wires.  We can in fact, without difficulty, include any finite number of ``ordinary'' creation-annihilation modes from each wire, thus including all degrees of freedom that overlap significantly with the junction under consideration.    These can be analyzed, as in Eq.\,(\ref{clifford}), into an even number of additional $\gamma$ operators, to include with the odd number of $b_j$.   But then the product $\Gamma^\prime$ of all these operators, including both types (and the appropriate power of $i$), retains the good properties Eq.\,(\ref{goodGammaProperties}) of the $\Gamma$ operator we had before.  

Now let us briefly discuss how $\Gamma$ resolves the puzzle in the previous section.  If $p \geq 5$, or even at $p=3$ with nearby electron interactions included, the emergent zero mode is a highly nonlinear entangled state involving all the wires that participate at the junction. The robustness of these conclusions results from the algebraic properties of $\Gamma$ we identified. 

%
%

{\it Pfaffian vortices.} ---It is interesting to compare the answer to a similar question in another physical context where Majorana modes arise \cite{ReadGreen}, that is, fractional quantum Hall effects of the Pfaffian type.   Following the notation and framework of \cite{nayakW}, appropriate wave functions for the state with four quasiparticles at positions $a, b, c, d$ can be constructed in the form
\begin{eqnarray}
\Psi_1 (&&z_j, a, b, c, d)   \nonumber \\
=&&~ {\rm Pf} \frac{(z_j - a)(z_j -b)(z_k - c) (z_k - d) + (j \leftrightarrow k)}{z_j - z_k} \ \Psi_0 (z_j), \nonumber \\
\Psi_2 (&&z_j, a, b, c, d) \nonumber \\
 =&&~ {\rm Pf} \frac{(z_j - a)(z_j -c)(z_k - b) (z_k - d)  +  (j \leftrightarrow k)}{z_j - z_k} \ \Psi_0 (z_j), \nonumber \\
\Psi_3 (&&z_j, a, b, c, d)\nonumber \\
 =&&~ {\rm Pf} \frac{(z_j - a)(z_j -d)(z_k - b) (z_k - c) + (j \leftrightarrow k)}{z_j - z_k} \ \Psi_0 (z_j), \nonumber \\
 {}
\end{eqnarray}
where ${\rm Pf}$ denotes the Pfaffian and $\Psi_0$ contains the standard Laughlin-Landau factors for filling fraction $1/2$.   Within the Pfaffian each quasiparticle acts on one member of a pair, and in each of $\Psi_1, \Psi_2, \Psi_3$ the quasiparticles themselves are paired off, so that each quasiparticle acts on the same electrons as its mate.  In $\Psi_1$ $ab$ and $cd$ are paired in this sense, and so forth.   It can be shown, by direct calculation, that $\Psi_1, \Psi_2, \Psi_3$ do not represent three independent states, since there is a ($a,b,c,d$-dependent) linear relation among them.   Two physical states remain.  This is the number required by a {\it minimal\/} implementation of the non-Abelian statistics, which can be based on the Clifford algebra with four generators \cite{ivanov}.      

Now formally we can take one of the quasiparticles off to infinity, and arrive at corresponding wave functions for three quasiparticles \cite{PfaffianFoot}:
\begin{eqnarray}\label{asymptoticPfaffians}
\tilde \Psi_1 (&&z_j, a, b, c)\nonumber\\  =&&~ {\rm Pf} \frac{(z_j - a)(z_j -b)(z_k - c)  + (j \leftrightarrow k)}{z_j - z_k} \ \Psi_0 (z_j), \nonumber \\
\tilde \Psi_2 (&&z_j, a, b, c) \nonumber\\ =&&~ {\rm Pf} \frac{(z_j - a)(z_j -c)(z_k - b)  + (j \leftrightarrow k)}{z_j - z_k} \ \Psi_0 (z_j), \nonumber \\
\tilde \Psi_3 (&&z_j, a, b, c) \nonumber\\ =&&~ {\rm Pf} \frac{(z_j - a)(z_k - b) (z_k - c) +  (j \leftrightarrow k)}{z_j - z_k} \ \Psi_0 (z_j). \nonumber \\
{} 
\end{eqnarray}
We find that there is no further reduction, so there is a two-dimensional space of states spanned by these wave functions, as required for a minimal (nontrivial) representation of the Clifford algebra with three generators.  In this context, then, it appears that the minimal spinor representation always suffices: no analogue of the electron parity operator is implemented.   

We conclude with the following comments.

(1) The algebraic structure defined by Eqs.\,(\ref{Psquared})-(\ref{Pb}) is fully nonperturbative.   It may be taken as the definition of the universality class supporting Majorana modes.  The construction of  $\Gamma$ (in its most general form) and its consequences [Eqs.\,(\ref{GammaSquared})-(\ref{GammaP})] reproduce that structure, allowing for additional interactions, with $\Gamma$ playing the role of an emergent $b$. The definition of $\Gamma$, the consequences Eqs.\,(\ref{GammaSquared}-\ref{GammaP}), and the deduction of doubling are likewise fully nonperturbative.  
(2) If we have a circuit with several junctions $j$, the emergent $\Gamma_j$ will obey the Clifford algebra 
\begin{equation}
\{ \Gamma_j, \Gamma_k \} ~=~ 2 \delta_{jk}.
\end{equation}
This applies also to junctions with $p =1$, i.e. simple terminals, and the circuit does not need to be connected.  
(3)$\Gamma$ is at the opposite extreme from a single-particle operator.   The corresponding mode is associated with the {\it product\/} wave function over the modes associated with the $b_j$.    In this sense we have extreme valence-bond (Heitler-London) as opposed to linear (Mulliken) orbitals.  The contrast is especially marked, of course, for large $p$. 
(4) The fact that interactions modify the Majorana modes in such a simply analyzed, yet highly nontrivial fashion, suggests new possibilities for circuit operations, which merit much further consideration.
(5) A Clifford algebra on an even number of generators that commute with the Hamiltonian can be reorganized, by inverting the procedure of Eq.\,(\ref{clifford}), into a supersymmetry algebra.   Thus our constructions support an emergent supersymmetry.   This supersymmetry algebra commutes with the Hamiltonian, but does not contain it. (Compare \cite{Qi2009}, where an emergent supersymmetry, relying on $T$ symmetry, has been discussed in the context of Majorana modes.)
(6) One can modify the preceding construction by using, in place of the $\Gamma_j$ matrices, matrices of the type
\begin{equation}
{\tilde \Gamma}_j ~\propto~ \sqrt H \, \Gamma_j
\end{equation}
to achieve a closed supersymmetry algebra, now including the Hamiltonian in the anticommutators.  One could also consider more elaborate construction, in which pieces of the total Hamiltonian are assigned to different $\Gamma_j$, exploiting locality conditions among the underlying $a$ operators to insure appropriate anticommutators. Of course the $\sqrt H$ operators themselves will not be local, except for specially crafted $H$. 

We thank Vincent Liu and Chetan Nayak for helpful comments. This work is supported by the U.S. Department of Energy under Contract No. DE-FG02-05ER41360.
J.L.  is supported in part by a Samsung Scholarship.



\begin{thebibliography}{99}
\bibitem{PfaffianFoot}{It is also possible to use wave functions based on the ``center-of-mass'' Pfaffians ${\rm Pf}  \frac{(z_j - a) (z_k -b ) (\frac{z_j + z_k}{2} -c ) + (j \leftrightarrow k) }{z_j-z_k}$, since either set can be expressed in terms of the other.
}
\end{thebibliography}
\end{document}